\documentclass[aps,twocolumn,superscriptaddress,nobibnotes,letterpaper]{revtex4-2}

\usepackage[pdftex]{graphicx}
\usepackage[pdftex]{epsfig}
\usepackage{amsmath}
\usepackage{amssymb}
\usepackage{amsfonts}
\usepackage{color}
\usepackage{wrapfig}
\usepackage{eucal}
\usepackage{hhline}
\usepackage{threeparttable}
\usepackage{supertabular}
\usepackage{multirow}
\usepackage{tabularx}
\usepackage{siunitx}
\usepackage{float}
\usepackage{sidecap}
\usepackage[version=4]{mhchem}

\usepackage{soul}

\usepackage{upgreek}

\usepackage{array}
\usepackage[colorlinks=true, allcolors=blue]{hyperref}

\newcommand{\und}[1]{_\textrm{#1}}

\definecolor{cgreen}{rgb}{.1,.6,.1}
\definecolor{co}{rgb}{.1,.6,.6}
\definecolor{orange}{rgb}{.9,.4,.0}

\newcolumntype{C}[1]{>{\centering\arraybackslash}p{#1}}

\begin{document}
\pagenumbering{arabic}

\title{An integrated microwave-to-optics interface for scalable quantum computing}

\author{Matthew J. Weaver}\thanks{These authors contributed equally to this work.}
\affiliation{QphoX B.V., Elektronicaweg 10, 2628XG Delft, The Netherlands}
\author{Pim Duivestein}\thanks{These authors contributed equally to this work.}
\affiliation{QphoX B.V., Elektronicaweg 10, 2628XG Delft, The Netherlands}
\author{Alexandra C. Bernasconi}\thanks{These authors contributed equally to this work.}
\affiliation{QphoX B.V., Elektronicaweg 10, 2628XG Delft, The Netherlands}
\author{Selim Scharmer}\thanks{These authors contributed equally to this work.}
\affiliation{QphoX B.V., Elektronicaweg 10, 2628XG Delft, The Netherlands}
\author{Mathilde Lemang}
\affiliation{QphoX B.V., Elektronicaweg 10, 2628XG Delft, The Netherlands}
\author{Thierry C. van Thiel}
\affiliation{QphoX B.V., Elektronicaweg 10, 2628XG Delft, The Netherlands}
\author{Frederick Hijazi}
\affiliation{QphoX B.V., Elektronicaweg 10, 2628XG Delft, The Netherlands}
\author{Bas Hensen}
\affiliation{QphoX B.V., Elektronicaweg 10, 2628XG Delft, The Netherlands}
\author{Simon Gr\"oblacher}\email{simon@qphox.eu}
\affiliation{QphoX B.V., Elektronicaweg 10, 2628XG Delft, The Netherlands}
\author{Robert Stockill}\email{rob@qphox.eu}
\affiliation{QphoX B.V., Elektronicaweg 10, 2628XG Delft, The Netherlands}


\begin{abstract}
Microwave-to-optics transduction is emerging as a vital technology for scaling quantum computers and quantum networks. To establish useful entanglement links between qubit processing units, several key conditions have to be simultaneously met:\ the transducer must add less than a single quantum of input referred noise and operate with high-efficiency, as well as large bandwidth and high repetition rate. Here we present a new design for an integrated transducer based on a planar superconducting resonator coupled to a silicon photonic cavity through a mechanical oscillator made of lithium niobate on silicon. We experimentally demonstrate its unique performance and potential for simultaneously realizing all of the above conditions, measuring added noise that is limited to a few photons, transduction efficiencies as high as 0.9\%, with a bandwidth of 14.8 MHz and a repetition rate of up to 100~kHz. Our device couples directly to a 50-$\Omega$ transmission line and can easily be scaled to a large number of transducers on a single chip, paving the way for distributed quantum computing.

\end{abstract}

\maketitle

Quantum processors and quantum networks have the potential to revolutionise information technology. They significantly expand on the existing capabilities of modern computers by solving previously intractable problems and enhancing the security of information sharing over networks~\cite{Kimble2008,Wehner2018,Alexeev2021}. In order to achieve superior performance compared to classical computers on practical computational tasks, state of the art quantum computers will need to increase the number of high-quality qubits by multiple orders of magnitude~\cite{Alexeev2021}. Several different realisations of such large scale quantum processors are currently being pursued~\cite{Ladd2010,DeLeon2021}, all with their own advantages and challenges. Some of the most advanced and promising quantum computers to date are based on superconducting qubits operating at cryogenic temperatures, which currently are capable of building processors with up to 127 qubits on a single chip~\cite{Chow2021}. However, engineering difficulties, such as the limited cooling power of cryogenic systems and space constraints of the microwave drive- and readout-lines pose severe restrictions on the scalability of these quantum computers.

Quantum interconnects can enable individual quantum modules to be incorporated into a multi-processor architecture, surpassing the limits of single cryogenic systems and chip scale fabrication~\cite{Awschalom2021,Krastanov2021,Bravyi2022}. At the same time, many leading stationary qubit implementations exchange information with microwave frequency photons limiting the reach of such connections to short distances in a cryogenic environment~\cite{Magnard2020}. Accordingly, transduction between microwave and optical frequencies is a key technology, enabling long-range channels at room temperature. Several different technologies are being explored for this task, including direct transduction~\cite{Mckenna2020,Xu2021,Sahu2022} and transduction assisted by an intermediary system such as a mechanical mode~\cite{Vainsencher2016,Higginbotham2018,Jiang2020,Mirhosseini2020,Stockill2021}, atomic state~\cite{Han2018,Fernandez-Gonzalvo2019,Bartholomew2020} or magnon mode~\cite{Hisatomi2016}. Despite these advances, a scalable quantum transducer which can interface with multiple qubits with high efficiency, large bandwidth and low added noise remains an outstanding challenge~\cite{Lauk2020,Han2021}.

\begin{figure*}
	\includegraphics[width=1.5\columnwidth]{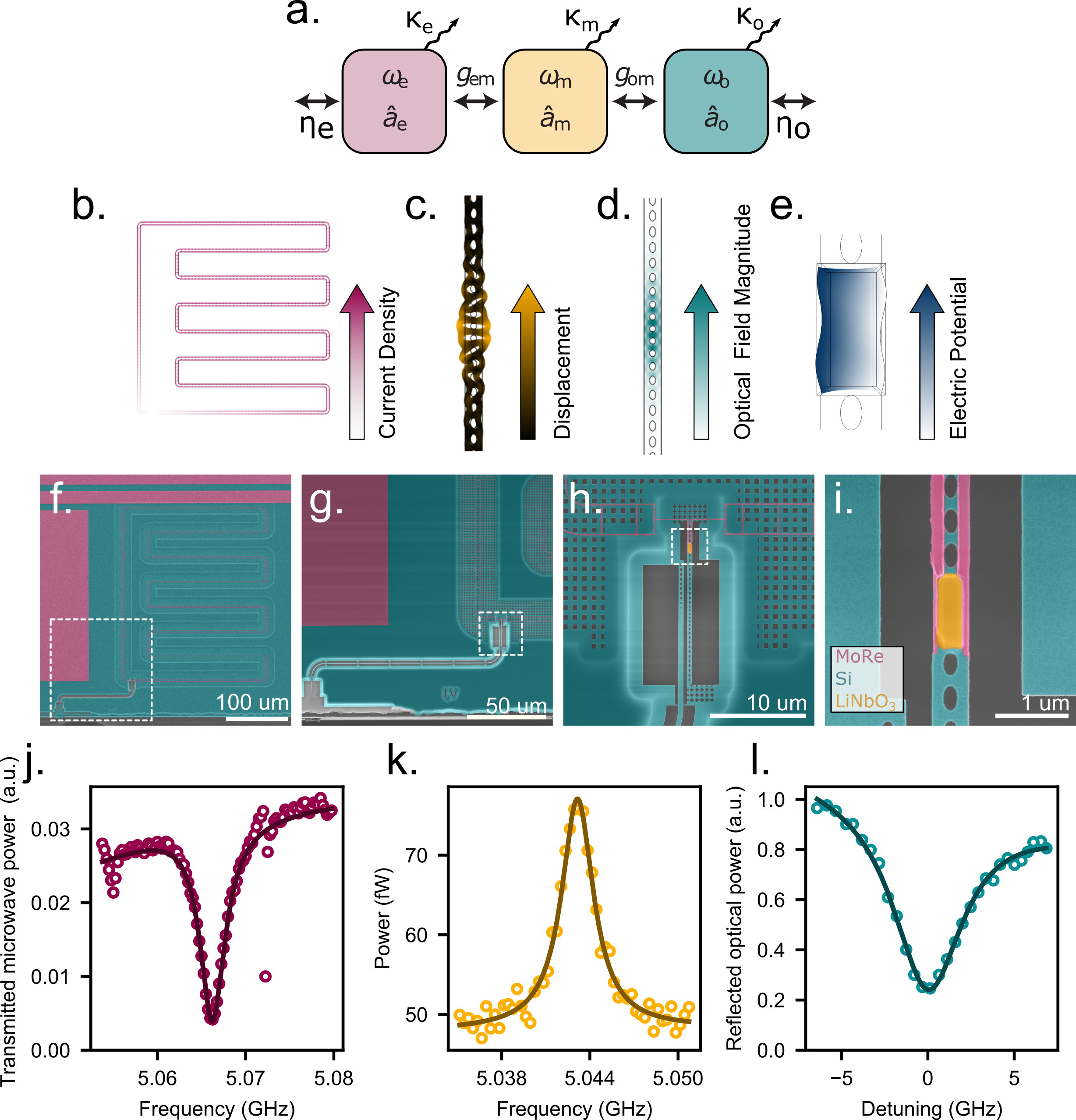}
	\caption{\textbf{Microwave-to-optical Transduction Device.} \textbf{a} The optomechanical transducer consists of three coupled bosonic modes:\ microwave (magenta), mechanical (yellow) and optical (cyan). Each has a frequency, $\omega/2\pi$, bosonic operator, $\hat{a}$, and loss rate, $\kappa/2\pi$.  The electrical and optical modes couple to additional external channels with efficiency $\eta$, as well as to the mechanical mode with coupling rate $g\und{em}$. \textbf{b-e} Finite element analysis of (\textbf{b}) the current density of the microwave resonator, (\textbf{c}) displacement of the mechanical mode, (\textbf{d}) field of the optical mode and (\textbf{e}) field of the piezo-electric resonator. \textbf{f-i} Scanning electron micrographs of the transducer. A superconducting molybdenum-rhenium nanowire loop microwave resonator (magenta) is capacitively coupled to an external transmission line (\textbf{f}). An optical waveguide separates the optical fiber output (bottom left) from the superconducting resonator (top right, \textbf{g}). The transduction region is fabricated into a silicon layer (cyan) and consists of a co-localized photonic and phononic crystal cavity linked by a phononic waveguide to a lithium niobate piezolectric block (yellow). The optical cavity is evanescently coupled to the optical waveguide (\textbf{h}). At the end of the phononic waveguide electrodes from the microwave resonator (magenta) contact the piezoelectric block (\textbf{i}). \textbf{j} Transmitted microwave power through the transmission line reveals the microwave resonance dip at 5.07~GHz. \textbf{k} The thermal motion of the mechanical resonance measured in the modulation of the optical signal reflected from the device. \textbf{l} The optical resonance of the photonic crystal cavity is measured using the reflected laser light from the transducer.}
	\label{Figure1}
\end{figure*}

The properties of the systems that would benefit from quantum transduction to optical wavelengths set the specific requirements for the transducer. Integrated two dimensional qubit systems have reached coherence times exceeding tens of microseconds~\cite{Siddiqi2021,Krinner2022}. To avoid a bottleneck, an ideal transducer should therefore operate with a repetition rate and bandwidth much greater than the inverse lifetime of the qubits ($\gg$10~kHz). Distributed quantum computation is likely to require many transducers in parallel to allow for fully remote gates between error corrected qubits~\cite{Horsman2012,Beals2013}, which necessitates an integrated system with many transducers on a single chip. Furthermore, most microwave frequency quantum systems can couple to transmission lines, which makes external coupling via 50-$\Omega$ lines an attractive and flexible approach~\cite{Krinner2019,Siddiqi2021}. Finally, the transducer must not adversely affect the performance of the quantum computation, requiring high efficiency and added noise below a single photon to generate remote entanglement between multiple processors~\cite{Zeuthen2020}. While devices based on 3-dimensional cavity geometries have demonstrated efficiencies up to 47\% with noise levels close to the quantum regime in one instance~\cite{Brubaker2022} as well as low added noise, $N\und{add}$, of 0.16 in another~\cite{Sahu2022}, these approaches are challenging to scale to large transducer numbers and require very high optical pump powers, which limits operation in cryogenic environments. These approaches typically also suffer from low bandwidth or repetition rates. Another chip scale approach recently demonstrated transduction with $N\und{add}\approx0.5$ directly coupled to a superconducting qubit~\cite{Mirhosseini2020}, but the quasiparticle generation from optical absorption destroyed the qubit state and limited the transducer repetition rate. So far no transducer operating in the quantum regime has demonstrated the bandwidth, repetition rate and scalability required for networking quantum processors.

Here we report on the experimental realisation of efficient, large bandwidth and low-noise transduction between microwave and optical frequencies by using a mechanical resonator as an intermediary. Our system is based on a thin-film lithium niobate (\ce{LiNbO3}) on silicon-on-insulator (SOI) platform~\cite{Jiang2022}, allowing us to simultaneously realise a strongly coupled electro-mechanical and a high-performance opto-mechanical interface. The device is extremely compact (0.15~mm$^2$, see Figure~\ref{Figure1}f) and scalable, converting between a 50-$\Omega$ impedance line and an optical fiber with a 14.8~MHz bandwidth, at repetition rates up to 100~kHz. We operate the transducer close to the quantum regime with only $6.2 \pm 1.8$ photons of added noise. Our chip design already features multiple transducers on a single chip and with modest improvements will allow for the generation of entanglement between remote quantum modules in parallel.

The central part of the transducer device consists of a suspended silicon nanobeam with a short phononic waveguide which is terminated by a piezoelectric block of lithium niobate. The phononic and photonic crystal of the nanobeam form a defect mode for both optical excitations around $\lambda\und{o}\approx1550$~nm and mechanical around $\omega\und{m}/2\pi\approx5$~GHz~\cite{Chan2012}. These modes are coupled to one another predominantly through the photoelastic effect. Owing to this interaction, application of an optical pump tone red-detuned from the optical resonance by the mechanical frequency realises a state swap between phonons occupying the mechanical mode and optical photons on resonance~\cite{Aspelmeyer2014}.

Interaction between the mechanical mode and microwave-frequency photons is realised through resonant piezoelectric coupling. In order to achieve this we include a small piece of lithium niobate at the end of the nanobeam, electrically contacted on each side (Fig.~\ref{Figure1}e). The holes of the nanobeam are shaped such that the mechanical mode leaks out of the defect and couples to the lithium niobate block, with the result that the different portions of the beam hybridise into a single extended mechanical mode.

\begin{figure*}
	\includegraphics[width=1.8\columnwidth]{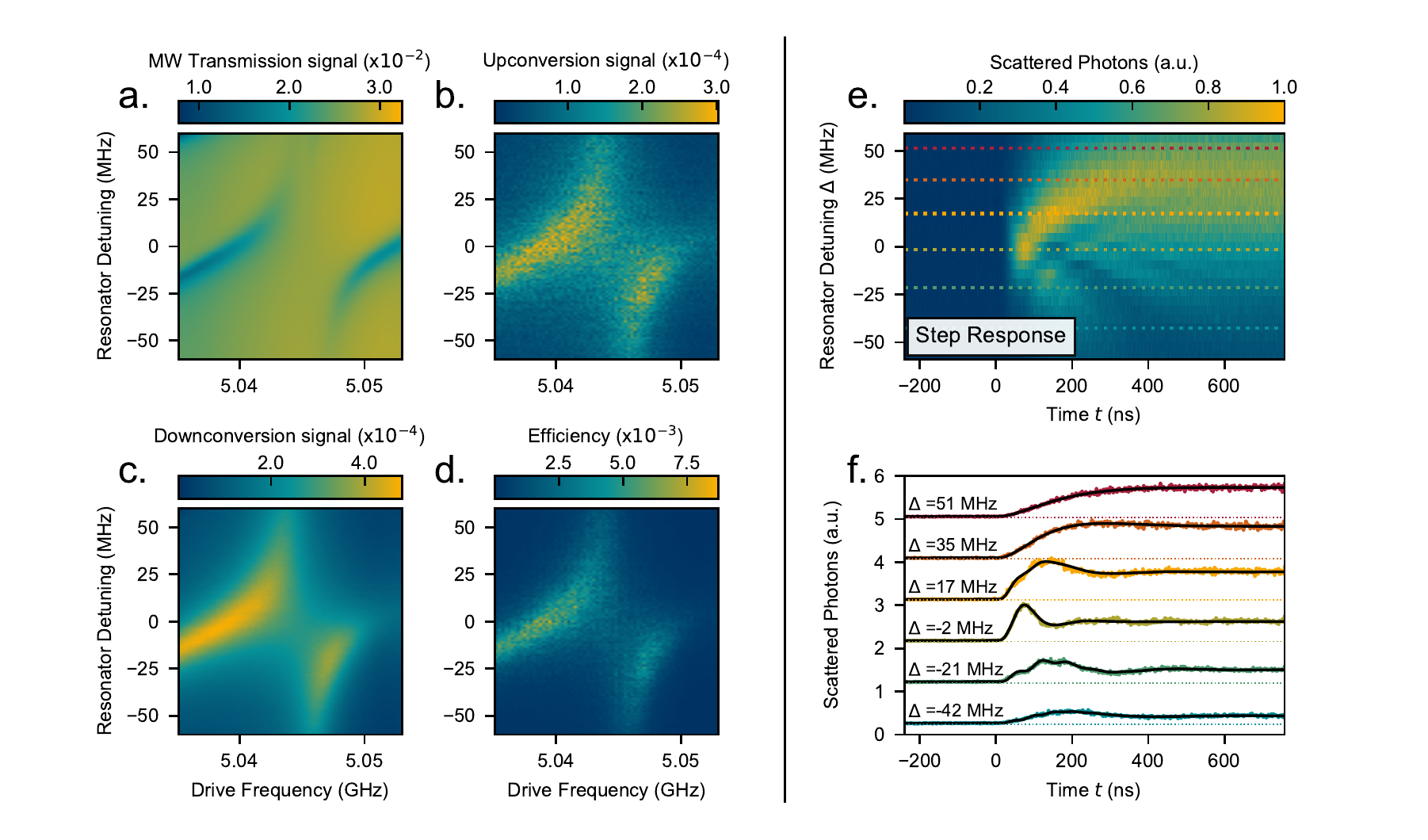}
	\caption{\textbf{Resonant Electromechanical Coupling and Bidirectional Transduction.} \textbf{a.} The frequency of the microwave resonator is tuned into resonance with the mechanical mode using a magnetic field in order to observe electromechanical coupling and transduction. We measure the microwave field going through the transmission line coupled to the microwave resonator. The avoided crossing at the frequency of the mechanical mode displays strong coupling between the mechanical and electrical mode with $g\und{em}/2\pi = 7.4 \pm$0.9~MHz \textbf{b.} Upconversion of the microwave signal to the optical domain detected using optical heterodyne measurement. \textbf{c.} Downconversion of an optical signal to the microwave domain collected via the microwave transmission line. As expected, both upconversion and downconversion follow the features of the hybridized electro-mechanical mode. \textbf{d.} The continuous drive transduction efficiency of the device is extracted using a 4-port measurement, revealing a maximum efficiency of $0.9\%$. \textbf{e} The step response of the mechanical resonance to a microwave drive which turns on at time 0. \textbf{f.} Indivdual cross-sections fit by a model yielding mean values of $g\und{em}/2\pi = 8.7 \pm$ 0.6~MHz  and $\kappa\und{e}= 4.9 \pm 0.5$~MHz .}
	\label{figure2}
\end{figure*}

In order to efficiently couple microwave photons between the 50-$\Omega$ transmission line and the high impedance piezoelectric device, we include a half-wavelength microwave resonator, terminated by the lithium niobate block (Fig.~\ref{Figure1}b,f). The resonator is formed from a 160-nm wide superconducting molybdenum-rhenium (\ce{MoRe}) nanowire. We pattern the resonator into a ladder geometry, which allows for tuning of the resonator frequency by up to 10\% by applying a magnetic flux resulting from a nearby superconducting coil~\cite{Xu2019}. This tuning range ensures that we can match the microwave resonator to the mechanical mode frequency of the device. Simulations of the resulting electrical, mechanical and optical modes of the full device predict simultaneous single-photon optomechanical coupling strengths of $g\und{om,0}/2\pi = 530$~kHz and piezoelectric coupling rates of up to $g\und{em}/2\pi = 18$~MHz. 

Our devices are fabricated from a 330~nm-thick layer of X-cut \ce{LiNbO3} bonded to a high-resistivity silicon-on-insulator wafer. We first pattern and remove the lithium niobate layer through ion-beam-milling, leaving the small block shown in Figure~\ref{Figure1}i. We then pattern and etch the silicon layer to form the nanobeam (Fig.~\ref{Figure1}h). Finally, we deposit the film of \ce{MoRe} for the resonator and suspend the device by selectively etching the oxide layer. We use etch holes around the microwave resonator to remove the oxide layer from underneath, increasing the microwave quality factor of the device.

The device is cooled down to around 25~mK inside a dilution refrigerator. A laser field is coupled through a lensed optical fiber into a tapered reflective waveguide, which is evanescently coupled to the optical resonator, allowing us to measure the optical mode in reflection through a circulator~\cite{Riedinger2016}. As a first step, we characterise the device parameters. Figure~\ref{Figure1}l shows the overcoupled optical resonance we measure when scanning the input laser frequency and recording the reflected optical power. We find the device's resonance at $\lambda\und{o}=1552.6$~nm, with an external coupling rate $\kappa\und{oe}/2\pi = 3.65\pm 0.14$~GHz and intrinsic loss rate $\kappa\und{oi}/2\pi = 1.34\pm 0.07$~GHz. We characterise the mechanical spectrum of the device by stabilising the laser frequency red-detuned by $\omega\und{m}$ with respect to the optical resonance and measuring the reflected signal through heterodyne detection. The resulting thermal spectrum of the mechanics is displayed in Figure~\ref{Figure1}k, showing a mode centred at $\omega\und{m}/2\pi=5.043$~GHz with a linewidth of $\kappa\und{m}/2\pi=2.63\pm0.12$~MHz. The microwave resonator on the chip is capacitively coupled to a 50-$\Omega$ transmission line. The microwave mode is measured in transmission and is displayed in Figure~\ref{Figure1}j, with a resonance centered at 5.07~GHz, an intrinsic loss-rate of $\kappa\und{ei}/2\pi=1.23\pm 0.11$~MHz and an external coupling rate of  $\kappa\und{ee}/2\pi=2.21\pm 0.10$~MHz. This measurement was taken with a magnetic field of approximately 3~mT (see S.I. for full table of parameters).

We now proceed to characterise the interaction between the mechanical and the microwave sub-components. The electromechanical coupling is measured by applying a continuous microwave drive tone and tuning the microwave resonator's frequency by applying a magnetic field (see Figure~\ref{figure2}a). As the resonator frequency approaches the optically-active mechanical mode at 5.043~GHz, we observe the characteristic avoided-crossing of the resonance frequency, as the mechanical and electrical modes become hybridised. From the size of this avoided crossing, we can directly extract the coupling rate between the mechanical mode and the microwave mode, which is $g\und{em}/2\pi = 7.4 \pm 0.9$~MHz, larger than $\kappa\und{e}$ and $\kappa\und{m}$. This strong electromechanical coupling is evidence of the high-cooperativity interface required for efficient and noise-free transduction. Given the mechanical and electrical damping rates, we extract a cooperativity of $C\und{em} = 4g\und{em}^2/\kappa\und{e}\kappa\und{m} = 24.2 \pm 4.4$.

In order to demonstrate the microwave-to-optics transduction capabilities of the device (upconversion), we input a frequency-swept microwave tone which we convert with a continuous optical pump red-detuned from the optical resonance. We measure the up-converted signal by detecting the resulting optical sideband in the reflected light from the device through heterodyne-detection. In order to measure transduction from optics to microwave (downconversion), we supply a signal on-resonance with the optical cavity by modulating the red-detuned pump using an electro-optic modulator, and directly measuring the output microwave tone from the device. The upconversion and downconversion scattering parameters (including all microwave and optical loss and gain in the setup) as a function of the microwave drive and the detuning of the microwave resonator are displayed in Figure~\ref{figure2}b and \ref{figure2}c, respectively. The transduction is at a maximum when the microwave-resonator approaches the mechanical frequency, and the two traces display the same avoided-crossing as in the microwave-spectroscopy.

We further supplement our continuous transduction measurements with measurements of the off-resonant reflected microwave and optical signals, which allow us to calibrate the loss and gain in the microwave and optical lines, and extract the bidirectional transduction efficiency of the transducer (see the Supplementary Information for details). We define transduction efficiency here as the waveguide to waveguide efficiency, excluding fiber coupling efficiency. The resulting 2-d map of efficiency is plotted in Figure~\ref{figure2}d. We find that, for a supplied optical pump power of 1~$\mathrm{\mu W}$ the peak efficiency we extract is around 0.9$\%$ and with a bandwidth of $2g\und{em} = 14.8 \pm 1.8$~MHz. For optical powers much greater than this, the efficiency is limited by non-linear effects such as thermal bistability of the optical cavity and reduction in the microwave-resonator quality factor due to residual absorption of the laser.

Pulsing the optical and microwave inputs allows us to access the time domain response of the device. We first look at the electromechanically enhanced excitation of the mechanical mode. In order to do this, we send in a continuous low-power optical drive and a fixed-frequency 2-$\mu s$-long microwave pulse on resonance with the mechanical mode. We filter the upconverted signal from the reflected red-detuned pump light with narrow-linewidth Fabry-P\'{e}rot resonators \cite{Riedinger2016}. The filtered light is detected using a superconducting nanowire single-photon detector (SNSPD). We correlate the arrival time of the photons following the start of the microwave pulse (at time $t$=0). By sweeping the detuning of the microwave resonator with respect to the microwave tone, $\Delta$, and measuring the scattered photons from the device, as shown in Figure~\ref{figure2}e,f, we can directly extract the time-dynamics of the transduction. As expected for a high-cooperativity interface, as the microwave resonator becomes resonant with the mechanical mode the rise time of the electro-mechanical loading is decreased and coherent oscillations between the resonator mode and the mechanical mode appear. We fit the time response to the microwave drive step with a simple model (see S.I.) to extract the electromechanical cooperativity of our device, which we find to be $C\und{em} = 33 \pm 10$, in good agreement with the expected value for our independently measured parameters.

\begin{figure}
\includegraphics[width=1\columnwidth]{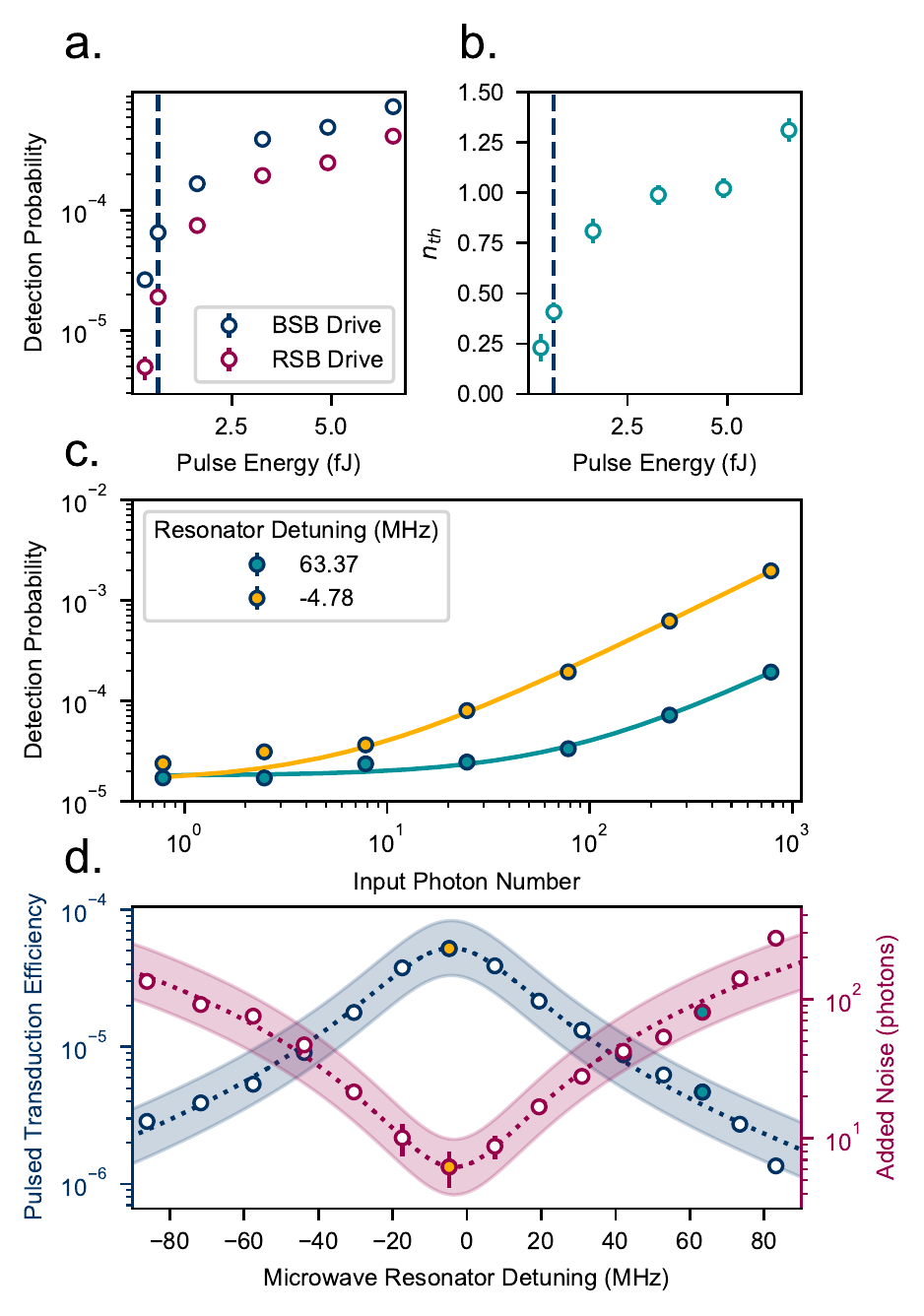}
\caption{\textbf{Pulsed Transduction and Added Noise.} \textbf{a} Detection probabilities for photons generated with an optical pump red detuned by $\omega\und{m}$(RSB) and blue detuned by $\omega\und{m}$(BSB) from cavity resonance, while the microwave drive is turned off. \textbf{b} The asymmetry between these two scattering rates can be used to extract the thermal occupation of the mechanical mode, $n\und{th}$, as a function of optical pulse energy. The dashed line indicates the optical pulse power used for the pulsed transduction measurements, a power for which the mechanical mode is predominantly in the ground state. \textbf{c} Detection probability of an upconverted optical photon as a function of input microwave photons for two different microwave resonator detunings. The linear fit allows us to extract the transduction efficiency and added noise. \textbf{d} Waveguide to waveguide transduction efficiency (blue) and added noise (magenta) as a function of microwave resonator detuning. The shaded regions indicate uncertainties arising from the microwave input attenuation. The yellow and green points correspond to the power sweeps in \textbf{c}. All error bars are one s.d.}
\label{figure3}
\end{figure}

Now that we have characterized the time dynamics of the electromechanical interface, we move to also using short optical pulses. We start by measuring the thermal occupation of the mechanical mode as a function of pulse energy using a sideband asymmetry measurement~\cite{Riedinger2016}. We use 40~ns long optical pump pulses, filtering out the reflected optical pump and sending the scattered photons at the cavity frequency to the SNSPD (total detection efficiency, $\eta\und{det}$ = 0.048). Figure~\ref{figure3}a shows the resulting detection probabilities and a clear asymmetry between blue and red detuned pulses. The extracted thermal occupation ($n\und{th}$) is shown in Figure~\ref{figure3}b. For low pulse energies $n\und{th}<1$, indicating that the mechanical mode is with a high probability in the quantum ground state. As microwaves are first loaded into the mechanical resonator with an efficiency $\eta\und{em}$ before thermal noise is added, the added noise referred to the input is $N\und{add}=n\und{th}/\eta\und{em}$ photons. For the subsequent pulsed transduction measurements we operate in the regime where the added thermal noise to the transduction process is limited to $N\und{add}\approx(0.41 \pm 0.05)/\eta\und{em}$ photons.

For a full pulsed transduction characterization of our device we now input 60~ns long microwave pulses fixed at the mechanical frequency into the transducer, followed by a 40~ns red-detuned optical pulse at a repetition rate of 100~kHz. The converted optical output photons from the transducer are filtered and detected using the SNSPD. Figure~\ref{figure3}c shows the photon detection probability as a function of input microwave photon number. The input microwave photon number is calibrated based on line attenuation measurements (see S.I.). The detected optical photons are the sum of the converted microwave photons ($\eta\und{det}\eta\und{om}\eta\und{em}n\und{mic}$) and the optomechanically converted noise photons ($\eta\und{det}\eta\und{om}N\und{add}$), where $n\und{mic}$ is the initial number of microwave photons, $\eta\und{det}$ the overall detection efficiency, $\eta\und{om}$ the optomechanical, and $\eta\und{em}$ the electromechanical transduction efficiency, respectively. We then sweep the detuning of the microwave resonator with respect to the mechanical frequency and record the pulsed transduction efficiency and the added noise, $N\und{add}$, shown in Figure~\ref{figure3}d. From this measurement we find a peak transduction efficiency of $(5.21 \pm 0.03)\times10^{-5}$ and minimum added noise of $N\und{add}=6.2\pm1.8$ photons.

We can use our model from the transduction time dynamics (shown in Figure~\ref{figure2}f) to determine the electromechanical loading efficiency. From the model we find an efficiency of $\eta\und{em}\sim0.12$, which is significantly higher than the $\eta\und{em} = 0.053\pm0.008$ extracted from the measurements in Figure \ref{figure3}d. From the microwave transmission data (see S.I.) we observe multiple strongly coupled modes which are relatively close in frequency to the mode of interest. By extending the model to include one extra neighboring mode, we see that the expected electromechanical loading efficiency drops to approximately 0.06 (see S.I.). With this reduced loading efficiency, the expected value for $N\und{add}$ from the sideband asymmetry measurement is $6.8\pm0.8$ photons, in good agreement with the calibrated value above.

\begin{figure}
\includegraphics[width=1\columnwidth]{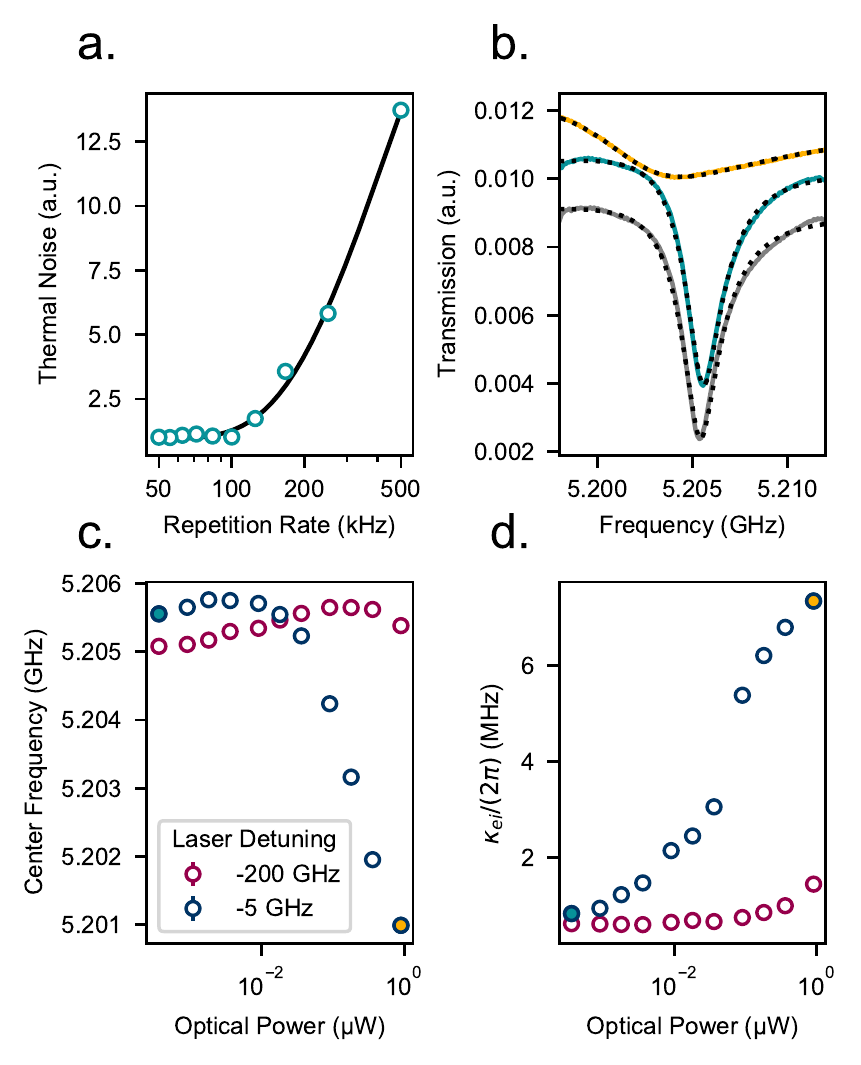}
\caption{\textbf{Repetition Rate and Quasiparticle Generation.} \textbf{a} The detected noise photons (normalised to 50~kHz repetition rate) increase with repetition rate. An exponential fit (solid line) yields the thermal recovery time. \textbf{b} We measure the microwave transmission through the resonator under three conditions: laser excitation off (bottom), pulsed excitation with 0.7 fJ pulses at 50 kHz repetition rate (middle) and continuous excitation with 1 $\mu$W (top). The microwave resonator frequency (\textbf{c}) and internal loss rate $\kappa\und{ei}$ (\textbf{d}) depend on the optical power sent to the transducer and the laser detuning from the optical resonance. The blue-filled (yellow-filled) points mark the powers and detunings used for the blue (yellow) traces shown in panel \textbf{b}.}
\label{figure4}
\end{figure}

As a final step we investigate the limits of our transduction rate. To do this, we input a red-detuned laser into the cavity and determine the thermal occupation of the mechanical mode by measuring optomechanically scattered noise photons, without a microwave drive. We perform this measurement for increasing repetition rate, which is shown in Figure~\ref{figure4}a. The relative thermal noise remains constant up to a repetition rate of approximately 100~kHz, above which, residual phonons from the previous excitation pulse increase the scattering rate. We find a thermal recovery time of 2~$\mathrm{\mu s}$, which is around the same order of magnitude as the mechanical-mode lifetimes in the system. 

We further measure the impact of optical light in the transducer on the frequency and quality factor of the superconducting resonator. Strong optical fields can pose a major challenge for integrated transducers as optical photons can generate quasiparticles (directly or via thermal heating) which degrade or even destroy the superconducting state~\cite{Mirhosseini2020}. In order to determine the impact of quasiparticle generation on our transducer, we measure the microwave resonance frequency shift and internal loss rate as a function of optical power (Fig.~\ref{figure4}c,d). Individual traces without light and with the optical powers and wavelengths used in the continuous (1-$\mu$W CW) and pulsed measurements (0.7-fJ pulses, 0.4-nW average power) are displayed in Figure~\ref{figure4}b. We see a frequency shift of $-3.74\pm0.07$~MHz and a  $9.6\pm1.0$ times greater internal loss rate in the continuous case, but in the pulsed case only a frequency shift of $+200\pm80$~kHz and a negligible loss rate change. We also repeat the optical power sweep 200-GHz detuned from the optical resonance and observe a drastically decreased response to optical power. We therefore conclude that the resonator changes are a result of the optical field inside the cavity, and that losses of the fiber-to-waveguide coupling play a negligible role. Overall, while we can observe some effects of quasiparticle generation, they are not significant enough to limit the low noise pulsed operation of our device.

Importantly, the bandwidth and repetition rates available with the transducer demonstrated here are already sufficient to interface with leading microwave-frequency quantum systems, such as superconducting qubits. The electromechanical efficiency for short microwave pulses can be further improved by optimizing the design of the piezoelectric interface to restrict coupling to near-resonant modes. Additionally, the noise contributions can be reduced with improved optical resonator linewidth, which is readily achievable through fabrication optimization;\ typical silicon-on-insulator optomechanical photonic crystal cavities have demonstrated intrinsic optical linewidths as low as 200 MHz~\cite{Zivari2022}. When these improvements are combined, our transducer enters the regime where $N\und{add}< 1$ and creating entanglement between remote qubits will be feasible.

We have realized a new generation of microwave-to-optics transducers, based on an integrated mechanical mode that serves as an intermediary between the two electromagnetic fields. Constructing the device from thin-film lithium niobate on silicon allows us to realize efficient, low-noise, and large bandwidth transduction, through scalable fabrication techniques. We measure transduction efficiencies of 0.9\% in the continuous and 5$\times10^{-5}$ in the pulsed regime, while adding only around 6 photons of input-referred noise.

The fully integrated device with a surface area of less than 0.15 mm$^2$ only requires 1~nW of average laser power, allowing for repetition rates of up to 100~kHz. In terms of area and power dissipation, the current transducer design therefore directly allows for scaling to more than 10,000 transducers in a single dilution refrigerator. With optical frequency multiplexing, these transducers could be used to interface a large scale quantum processor using only around 100 fibers (with a predicted passive heatload of 300 pW \cite{Lecocq2021}). Our device is also impedance matched to a 50-$\Omega$ line, allowing flexible connections to qubits located within the same dilution refrigerator. This enables complete separation between the transduction and the qubit chip.

Our system is ideally suited to operate in the few-photon regime and optically read-out the state of a superconducting qubit, as has been recently demonstrated in another system~\cite{Delaney2022}. Here the requirements on added noise and efficiency are relaxed and our current performance is sufficient to perform single-shot readout via the readout resonator of a qubit. Combined with optical multiplexing, this could lead to a system that replaces superconducting microwave amplifiers and cables, removing significant heatload and space restrictions inside a dilution refrigerator, enabling quantum processors to scale.

\begin{acknowledgments}
We would like to thank Amir Safavi-Naeini for valuable discussions and support. We thank Jules van Owen and Damaz de Jong at Qblox for technical support. We further gratefully acknowledge assistance from Emma He, helpful discussions with Christophe Jurczak and Johannes Fink, supply of wafers from NGK Insulators Ltd.\ and the hospitality of the Department of Quantum Nanoscience at Delft University of Technology, as well as the Kavli Nanolab Delft. This work is financially supported by the European Innovation Council (EIC Accelerator QModem 190109269), the Province of Zuid-Holland (R\&D samenwerkingsproject QConnect).
\end{acknowledgments}

\section*{Conflicts of interest}
All authors declare a financial interest in QphoX B.V.

\section*{Data Availability}
Source data for the figures will be made available on Zenodo.

\setcounter{figure}{0}
\renewcommand{\thefigure}{S\arabic{figure}}
\setcounter{equation}{0}
\renewcommand{\theequation}{S\arabic{equation}}

\clearpage

\onecolumngrid

\begin{center}
	{\Large \textbf{Supplementary Information}}
\end{center}

\vspace{1cm}

\twocolumngrid

\section{Fabrication Details}

The starting material for these devices is a thin film of X-cut LiNbO$_3$ (LN) bonded on a high resistivity silicon-on-insulator (SOI) substrate. The full device fabrication consists of four main steps. First, an Ar milling process patterns the LN layer to form small blocks. Second, a reactive-ion etch (RIE) step forms the nanobeams. Third, deposition of a MoRe layer and lift-off creates the bondpads, the feedline, and the resonators. Finally, an oxide etch selectively removes the sacrificial SiO$_2$ layer underneath the Si device layer to suspend the devices. This step also removes the oxide under the resonators, increasing the electrical quality factor of the devices.

\section{Experimental Details}

\begin{figure*}
	\includegraphics[width=1.4\columnwidth]{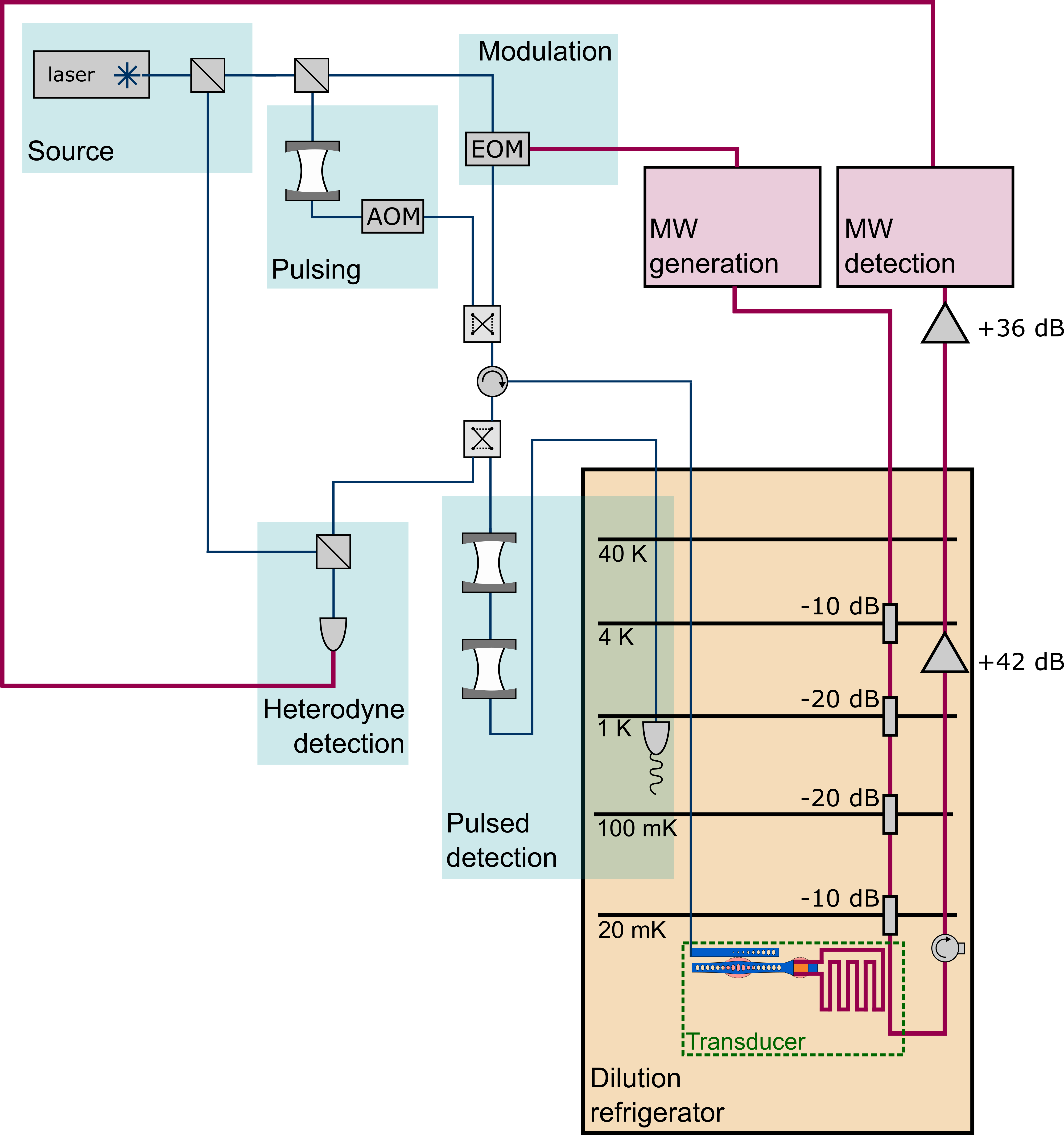}
	\caption{\textbf{Experimental Setup} Our experimental setup consists of optical and microwave inputs and outputs to the transducer. The transducer is mounted on the mixing chamber plate of a dilution refrigerator. The optical input path passes through and amplitude electro-optic modulator (EOM) or a filter cavity and an acousto-optic modulator (AOM). The output path leads to either heterodyne detection or two filter cavities followed by a superconducting nanowire single photon detector (SNSPD). }
	\label{SI_setup}
\end{figure*}

\begin{table*}
	\begin{tabular}{c | l | c | c | c}
		Parameter name & Parameter Description & Value & Error & Units \\
		\hline
		Device Property & & & \\
		\hline
		$\omega\und{m}/2\pi$ & Mechanical Frequency & 5.043 & & GHz\\
		$\omega\und{o}/2\pi$ & Optical Frequency & 193.087 &  & THz\\
		$\kappa\und{m}/2\pi$ & Mechanical Linewidth & 2.63 & 0.12 & MHz\\
		$\kappa\und{e}/2\pi$ & Electrical Linewidth & 3.439 & 0.144 & MHz\\
		$\kappa\und{ee}/2\pi$ & External Electrical Coupling Rate & 2.213 & 0.097 & MHz\\
		$\kappa\und{ei}/2\pi$ & Intrinsic Electrical Loss Rate & 1.226 & 0.106 & MHz\\
		$\kappa\und{o}/2\pi$ & Optical Linewidth & 4.99 & 0.19 &GHz\\
		$\kappa\und{oe}/2\pi$ & External Optical Coupling Rate & 3.65 & 0.14 &GHz\\
		$\kappa\und{oi}/2\pi$ & Intrinsic Optical Loss Rate & 1.34 & 0.07 &GHz\\
		$\eta_e \equiv \frac{\kappa\und{ee}}{2\kappa\und{e}}$ & Electrical Resonator Coupling & 0.322 & 0.019 & \\
		$\eta_o \equiv \frac{\kappa\und{oe}}{\kappa\und{o}}$ & Optical Resonator Coupling & 0.731 & 0.040  &\\
		$g\und{om}/2\pi$ & Single Photon Optomechanical Coupling Rate & 561  & 51 & kHz\\
		$g\und{em}/2\pi$ & Electromechanical Coupling Rate & 7.4 & 0.9 & MHz\\
		\hline
		Experimental Setup & & & \\
		\hline
		$\eta\und{fiber}$ & Fiber Coupling Efficiency & 0.42 & 0.03 & \\
		$\eta\und{filter}$ & Filter Transmission Efficiency & 0.19 & 0.03 & \\
		$\eta\und{SNSPD}$ &  SNSPD Detection Efficiency & 0.60 & 0.03 & \\
		$\eta\und{MW}$ & Line attenuation fridge and external & -113.6 & 0.2 & dB\\
	\end{tabular}
	\caption{Assembled parameters for the transducer and experimental setup in the main text.}
	\label{SI_parameters}
\end{table*}

Our experimental setup consists of a microwave and an optical interface for the transducer as shown in Figure~\ref{SI_setup}. The transducer is located on the baseplate of a dilution refrigerator at a temperature of around 25 mK. Microwave lines with attenuators deliver the microwave input signals to the baseplate and then to the transmission line on the sample. The output of the on-chip transmission line is connected to two circulators and then to a high-electron-mobility transistor (HEMT) amplifier at 4K and a room temperature amplifier. Microwave signals are generated with a microwave source with a fast microwave switch or a vector network analyser (VNA) and detected with a VNA.

The optical input path consists of a pulsed path and a continuous path. For continuous mode transduction we lock the laser red-detuned from the optical cavity by $\omega\und{m}$ and pass it through an amplitude electro-optic modulator (EOM). The optical input signal for downconversion is generated by driving the EOM at $\omega_m$ and generating a sideband at the cavity frequency. For pulsed transduction, we generate pulses with a function generator driving an acousto-optic modulator (AOM). The optical input is passed through a fiber into the cryostat and on the baseplate we couple the light into the on chip waveguide of the transducer with a lensed fiber tip.

We seperate the reflected output optical signal from the input signal with a circulator. For continuous mode transduction we detect the output signal with heterodyne detection using a high bandwidth photodiode. For pulsed transduction we pass the output signal through two filter cavities to remove the optical pump. We then pass the signal back into the refrigerator for detection and photon counting on a single photon detector (SNSPD)~\cite{Riedinger2016}. The total detection efficiency is $\eta\und{det} = \eta\und{fiber}\eta\und{filter}\eta\und{SNSPD}$ = 4.8\%.

For continuous mode transduction we determine the transduction efficiency with a four port VNA measurement using the method developed by Andrews et al.~\cite{Andrews2014}. By measuring the microwave attenuation in the microwave input and output ($S\und{ee}$) and the microwave attenuation in the optical input and output ($S\und{oo}$), we can calibrate the input and output losses, which are included in the upconversion ($S\und{oe}$) and downconversion ($S\und{eo}$) measurements. The EOM generates two sidebands; both contribute to $S\und{oo}$, but only the sideband resonant with the cavity contributes to $S\und{eo}$. Therefore the continuous transduction efficiency, $\eta\und{CW}$, is given by:

\begin{equation}
	\eta\und{CW} = \frac{2\alpha \left|S\und{oe}\right|\left|S\und{eo}\right|}{\left|S\und{ee}\right|\left|S\und{oo}\right|},
\end{equation}

where the pre-factor $\alpha = 0.73$ corresponds to the reduction in $S\und{oo}$ we measure for a carrier red-detuned by 5-GHz compared to a fully off-resonant tone. Because the method cancels all common paths, the bidirectional continuous transduction efficiency we measure here is from one port of the on-chip microwave waveguide to the on-chip optical waveguide.

In the strong coupling electro-mechanical regime, the measurement of the transduction bandwidth is complicated as the efficiency no longer follows a Lorentzian lineshape from which we can extract a full width at half maximum (FWHM). However, high transduction efficiency is maintained across the whole avoided crossing with a separation given by 2$g\und{em}$. Therefore we use 2$g\und{em}/2\pi = 14.8 \pm 1.8$~MHz as the transduction bandwidth in the main text. Increasing $\kappa\und{ee}$ further by reducing the gap between the resonator and the feedline woudl 

Measured device and setup parameters are included in Table~\ref{SI_parameters}. All parameters are measured for the device described in the main text, apart from the value of $g\und{om}$. We extract this parameter from the scattering probability in the pulsed experiment shown in Figure 3 a-b, following the same method as in \cite{Hong2017}.

\section{Microwave Input Loss Calibration}

For continuous wave transduction we can use a four port measurement to remove external setup losses from the transduction efficiency. However, in the case of pulsed microwave to optics transduction, we need to calibrate the input microwave losses to the transducer to determine the efficiency and added noise. Because microwave losses are temperature dependent, directly measuring losses through the cables at room temperature gives only an estimate of the input microwave loss. In this section we describe an alternative measurement of the input microwave loss.

We use a three line measurement to directly measure the losses. The three lines are the input microwave line which drives the transducer and two additional microwave lines with no added attenuation, which are nominally identicle. We connect the three lines via a 3-port circulator. We measure the transmission from one unattenuated line to the other, subtract circulator losses and divide by 2 to determine an unattenuated line loss of $6.56 \pm 0.03$~dB. We then measure the losses going through the drive line and one attenuated line and subtract circulator losses and the unattenuated line loss, arriving at a microwave drive line attenuation of $69.97 \pm 0.18$~dB. 

The connection between the microwave drive line and the device is more uncertain. The connecting wire between the drive line and the device has an attenuation of $2.74 \pm 0.02$ dB at room temperature. However, typically attenuation decreases at low temperatures and we must add any chip connector losses. Given these systematic uncertainties we estimate a loss for this segment of $2 \pm 2$~dB. $N\und{add}$ and transduction efficiency in Figure 3c of the main text show a shaded region corresponding to a line attenuation of $71.97 \pm 2$dB.

To measure the pulsed transduction efficiency we calibrate the input microwave photons with the sum of the measured dilution refrigerator and external line attenuation ($\eta\und{MW}$). The pulsed transduction efficiency, $\eta\und{pulsed}$ is defined as:
\begin{equation}
	\eta\und{pulsed} = \frac{n\und{det}\hbar\omega\und{drive}}{\eta\und{MW}\eta\und{det}P\und{pulse}t\und{pulse}}
\end{equation}
$n\und{det}$ is the mean number of photons detected on the detector per excitation pulse. $P\und{pulse}$, $t\und{pulse}$ and $\omega\und{drive}$ are the microwave power, duration and frequency of the excitation pulse at the output of the microwave generator. The pulsed transduction efficiency is the on-chip microwave waveguide to on-chip optical waveguide transduction efficiency, as in the continuous case.

\section{Microwave Resonator Tuning and Coupling}

\begin{figure}
	\includegraphics[width=\columnwidth]{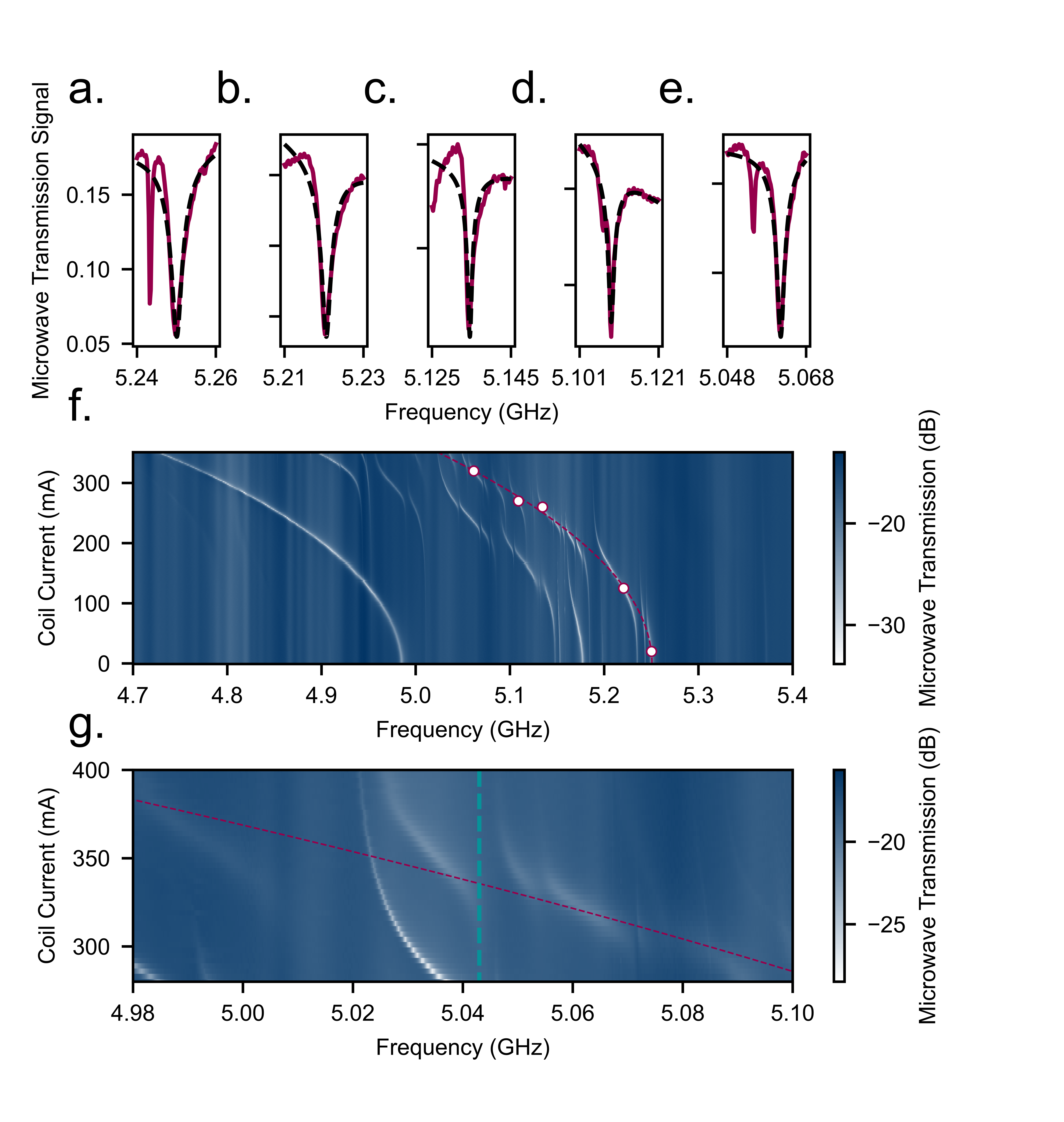}
	\caption{\textbf{Microwave resonator tuning} \textbf{f.} We scan the microwave transmission through the transmission line for a range of coil currents applied to an external coil. The transmission line is connected to three microwave resonators, which all tune in frequency with the applied field. The lowest frequency resonator is patterned without a LN block. In \textbf{a.} to \textbf{e.} we plot and fit individual cross-sections of the microwave resonator used in the main text. We can extract a tuning rate of 1.8~GHz$\mathrm{A}^{-2}$. \textbf{g.} We zoom into the region around the transduction mode (frequency shown by green line) to show other modes close in frequency. The measurement in this subfigure was taken with 1 $\upmu$W of optical power, which broadens the microwave response. The bright line between 5.04 GHz and 5.02 GHz is a feature from a different microwave resonator.}
	\label{SI_figure1}
\end{figure}

Our microwave resonators are laddered kinetic inductance resonators similar to those implemented by Xu et al. \cite{Xu2019}. By including loops in the wire, the kinetic inductance can be increased with the application of a magnetic field, tuning the frequency. We mount a coil close to the sample to generate the magnetic field. Figure \ref{SI_figure1} displays magnetic field tuning for three microwave resonators coupled to a single transmission line. We fit the resonance frequency of the resonator used in the main text as a function of coil current (Figure \ref{SI_figure1}a-e) and extract a tuning rate of 1.8~GHz$\mathrm{A}^{-2}$ (Figure \ref{SI_figure1}f). We estimate the magnetic field at the resonator to be $\sim10$ mTA$^-1$. In total, we can achieve up to 500 MHz of tuning.

In Figure \ref{SI_figure1}g, we also scan the microwave transmission in a narrower range around the transduction frequency. There are a number of avoided crossings close to the transduction frequency, corresponding to other mechanical modes of the system.

\section{Electromechanical Theoretical Model}

We use a simplification of the model outlined by Han et al.~\cite{Han2021} to determine the time dynamics of energy transfer between modes. This model can be used to describe all mechanical, electrical and optical modes in the system. In our specific case, we operate in the regime where $C\und{om}\ll1$. In this regime the optical emission from the transducer is directly proportional to the mechanical occupation of the transduction mode and both the optical driving field and the occupation of the optical cavity mode have negligible effect on the occupation of the mechanical resonator. 

The model involves a first order differential equation for the time evolution~\cite{Han2021}:
\begin{widetext}
	\begin{equation}
		\begin{split}
			\frac{d}{dt}\left( {\begin{array}{c}
					a\und{e}  \\
					a\und{m}  \\
			\end{array} } \right)
			= \left( {\begin{array}{cc}
					-i\Delta\und{e} - \kappa\und{e}/2   & ig\und{em}   \\
					ig\und{em}  &  -i\Delta\und{m} - \kappa\und{m}/2 \\   
			\end{array} } \right)
			\left( {\begin{array}{c}
					a\und{e}  \\
					a\und{m}  \\
			\end{array} } \right)
			+ \left( {\begin{array}{c}
					\sqrt{\kappa\und{ee}/2}c\und{in}  \\
					0  \\
			\end{array} } \right)
		\end{split}
	\end{equation}
\end{widetext}

$a\und{e}$ and $a\und{m}$ are the microwave and mechanical mode occupancies. $\Delta\und{e}$ and $\Delta\und{m}$ are the detuning of the microwave resonator and the mechanical mode from the drive frequency respectively and $c\und{in}$ is a parameter proportional to the microwave power sent into the transmission line. We can analytically solve this equation and find the time dynamics of the mode populations as a function of $\Delta\und{e}$, $g\und{em}$, $\kappa\und{e}$ and time. We numerically fit the function of $\Delta\und{e}$, $g\und{em}$ and $\kappa\und{e}$ to the time dynamics in Figure 2e,f of the main text and extract the parameters $g\und{em}$ and $\kappa\und{e}$. The model succesfully explains the key features of electromechanical loading with long, 2 $\upmu$s pulses.

\begin{figure}
	\includegraphics[width=1\columnwidth]{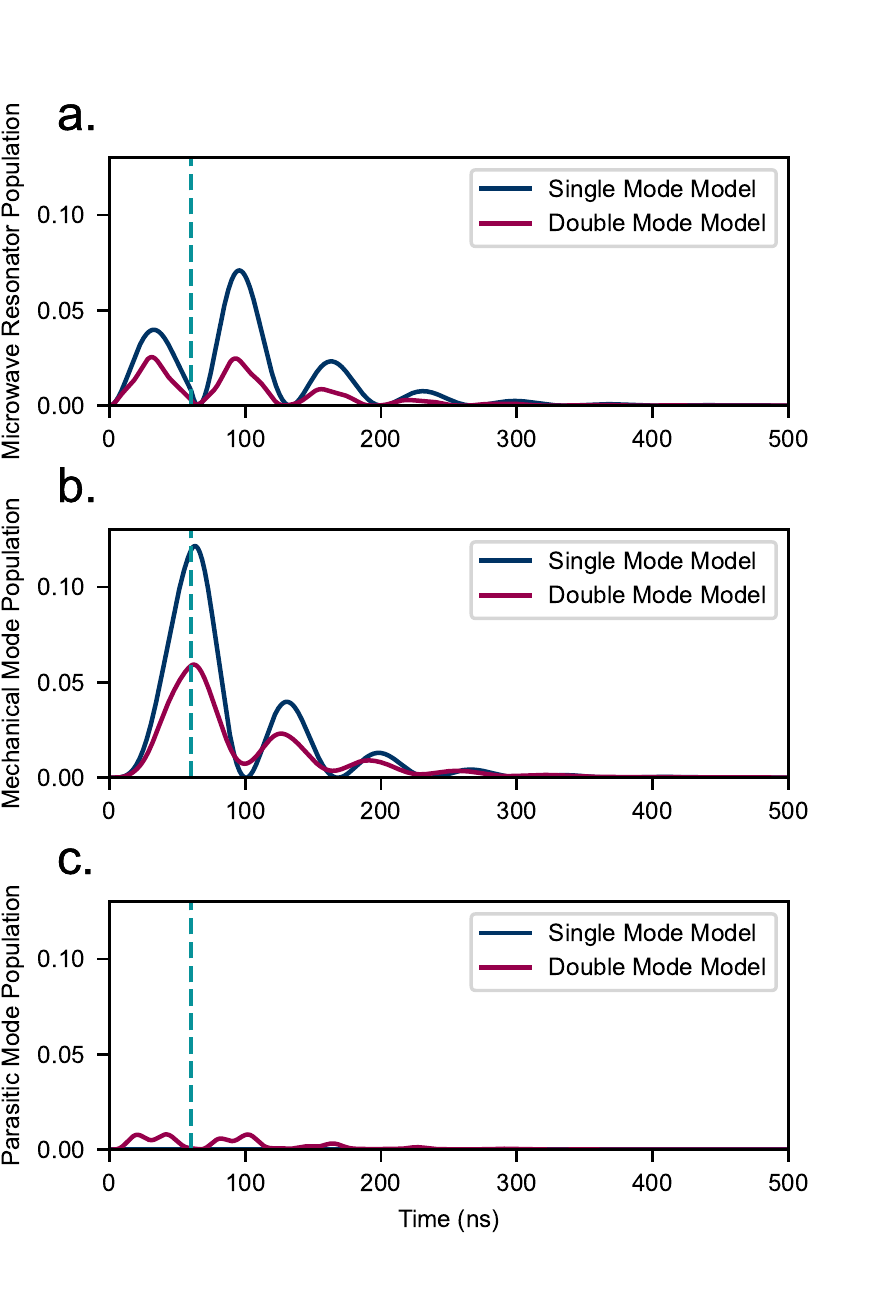}
	\caption{\textbf{Short pulse loading dynamics} We study the time response to a 60 ns, one photon microwave pulse of the mode occupation of the (\textbf{a}) microwave mode, (\textbf{b}) main mechanical mode and (\textbf{c}) parisitic extra mechanical mode. In the simpler single mode model only (\textbf{a}) and (\textbf{b}) are considered, but we also consider a double mode model with all three modes. In the extended model the transduction efficiency is reduced.}
	\label{SI_figure2}
\end{figure}

We also investigate the loading efficiency with the short 60 ns microwave pulses we used in Figure 3 of the main text. We plot the mode occupation of the microwave resonator and mechanical mode when the systen is driven with a single microwave photon in Figure~\ref{SI_figure2}. Energy is exchanged between the two modes as expected for the strong coupling of our device. We reach a peak phonon population or loading efficiency of 0.12, which is greater than the measured value for loading efficiency. This motivates us to investigate some of the other mechanical modes in the system (see Figure~\ref{SI_figure1}g).

We extend the model to include an extra parasitic mechanical mode:

\begin{widetext}
	\begin{equation}
		\begin{split}
			\frac{d}{dt}\left( {\begin{array}{c}
					a\und{e}  \\
					a\und{m}  \\
					a\und{m2} \\
			\end{array} } \right)
			= \left( {\begin{array}{ccc}
					-i\Delta\und{e} - \kappa\und{e}/2   & ig\und{em}  & ig\und{em2} \\
					ig\und{em}  &  -i\Delta\und{m} - \kappa\und{m}/2 & 0\\   
					ig\und{em2}  &  0 & -i\Delta\und{m2} - \kappa\und{m2}/2 \\   
			\end{array} } \right)
			\left( {\begin{array}{c}
					a\und{e}  \\
					a\und{m}  \\
					a\und{m2} \\
			\end{array} } \right)
			+ \left( {\begin{array}{c}
					\sqrt{\kappa\und{ee}/2}c\und{in}  \\
					0  \\
					0  \\
			\end{array} } \right)
		\end{split}
	\end{equation}
\end{widetext}

$a\und{m2}, g\und{em2}, \Delta\und{m2}$ and $\kappa\und{m2}$ are the second mode population, the electromechanical coupling rate for the second mode, the detuning of the drive from the second mode and the mechanical loss rate of the second mode. We plot the time response for the above derived parameters, $g\und{em2}/2\pi$ = 17~MHz, $\Delta\und{m2}/2\pi$ = 28~MHz and $\kappa\und{m2} = \kappa\und{m}$, the approximate parameters of the large avoided crossing closest below the frequency of the main mode in Figure~\ref{SI_figure2}. We did not measure $\kappa\und{m2}$, but the exact value does not have a large impact on the results.

We see a large decrease in the mode occupation of both the microwave and mechanical modes in the extended double mode model compared to the single mode model. In particular the maximum phonon occupation drops to 0.06. This leads to a decrease in the electromechanical loading efficiency. If we put in the parameters for other neighboring mechanical modes, the maximum phonon occupation also drops, but only by 5\%. Determining the exact mode couplings and loss rates is left to a future investigation.

For long pulses (for example 2 $\upmu$s as used in Figure 2 e,f of the main text) and continuous driving, we see negligible differences between the single mode and double mode model results. The insensitivity to other mechanical modes is caused by the reduced bandwidth of longer pulses compared to the 60 ns pulses. With a reduced bandwidth for the drive, the detuning is much larger than the bandwidth resulting in only a small interaction.

\section{Sideband Asymmetry Power Correction}

\begin{figure}
	\includegraphics[width=1\columnwidth]{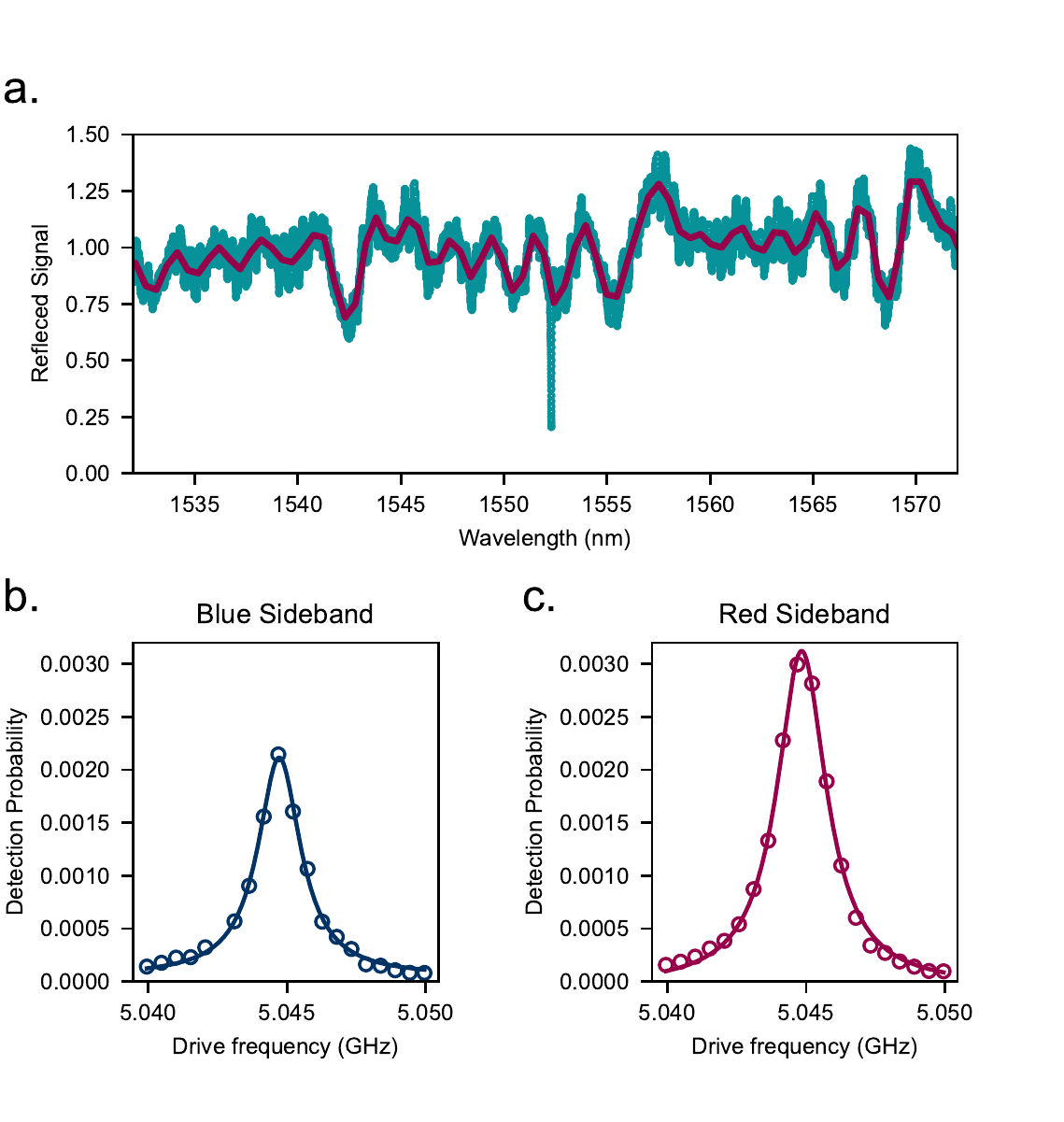}
	\caption{\textbf{Sideband Asymmetry Calibration} \textbf{a.} A wavelength scan of the optical reflection of the device displays periodic oscillations due to reflections in the optical waveguide(green). We normalize the signal to the mean and plot a smoothed trend line(magenta). In order to calibrate for optical power in the cavity for different wavelengths, we drive the mechanical mode to a large occupation and measure the detected photons from a optical pulses detuned by $\omega\und{m}$ to the blue (\textbf{b}) or the red (\textbf{c}). We derive a correction factor for input power from the ratio between the two fitted peaks.}
	\label{SI_figure3}
\end{figure}

Our devices are fabricated with a 150 $\upmu$m long extended optical waveguide to prevent stray light from the coupling lensed fiber directly impacting the microwave resonator. Reflections at the beginning (and to a lesser extent at the supporting tethers) of the waveguide generate Fabry-Perot type interferences in the reflection from the waveguide. This is visible in the oscillations of the reflected signal of the long wavelength scan of Figure \ref{SI_figure3}a. We see that the much deeper and narrower reflection dip from the optical cavity is superimposed on this oscillation.

Although the reflections that cause these oscillations are small, they also affect the optical power delivered to the microwave cavity as a function of detuning. We measure the relative optical cavity occupation with the pump laser detuned red and blue by the mechanical frequency by exciting the mechanical transduction mode to a large occupation with a microwave drive via the transmission line and detuned microwave resonator. For the large estimated mechanical occupations of order 100 phonons, the optomechanical scattering rate differences between red and blue sideband should be negligible. We measure, however in Figure \ref{SI_figure3}b and c, that the detection probability of photons from the two sidebands is quite different, which we directly attribute to the wavelength dependent interferences in the waveguide.

In the main text we measure the asymmetry in detection rates as a means to extract the thermal occupation of the mechanical mode. This technique assumes equal optical cavity occupation for both sidebands. Therefore in order to satisfy this assumption, we adjust the optical input power we send to the device by increasing the blue detuned power by a factor of 1.53, the ratio of the fitted peaks in Figure \ref{SI_figure3}. Repeating this calibration with the other optomechanical mode at 5.072 GHz or with continuous optical driving and heterodyne detection gives a similar correction factor, indicating that it is not mechanical mode or collection path dependent.

\section{Other transduction mode}

\begin{figure}
	\includegraphics[width=1\columnwidth]{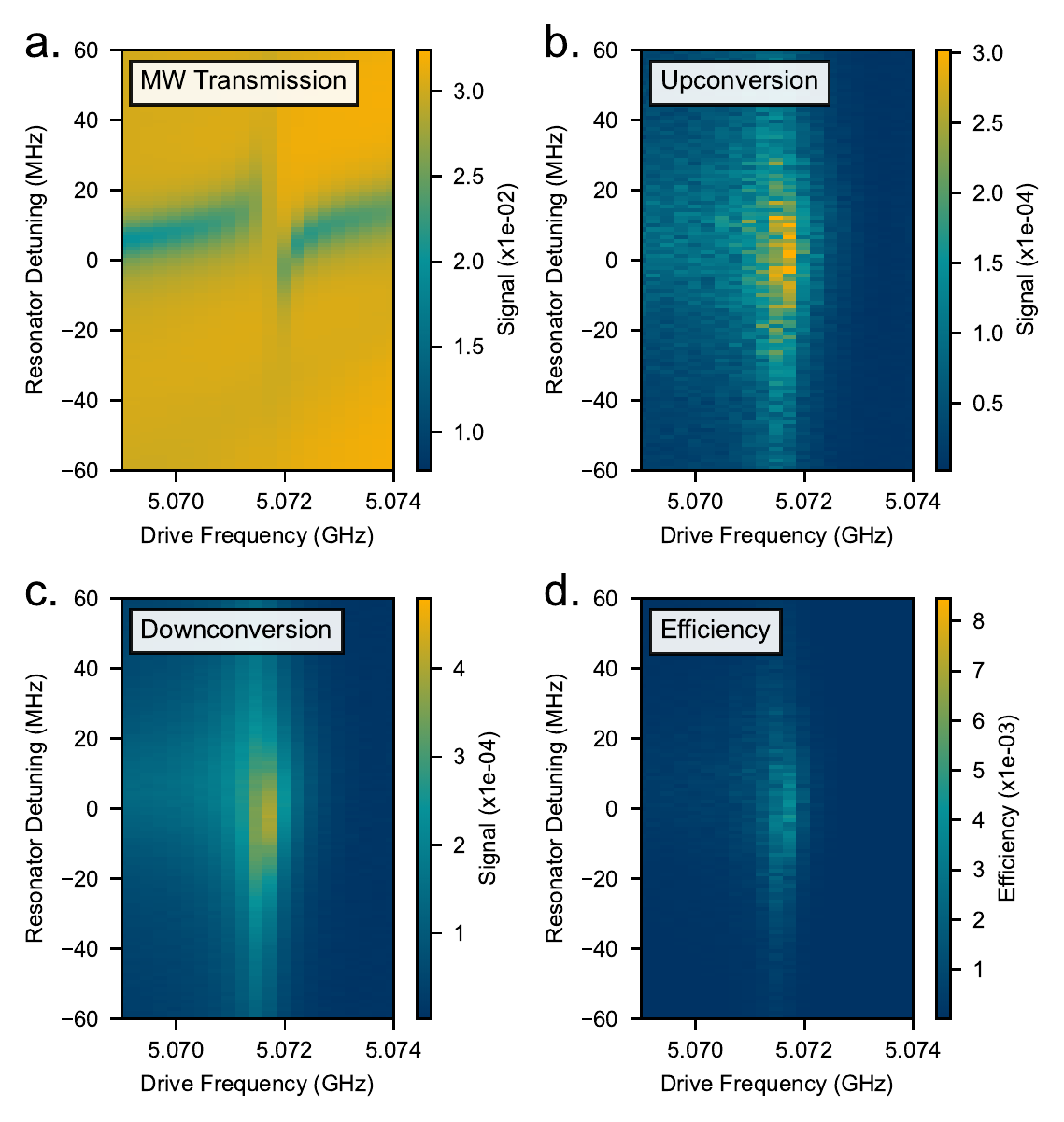}
	\caption{\textbf{Continuous Transduction on another mode} We remeasure the continuous wave transduction on a different mechanical mode. \textbf{a.} The microwave transmission shows an avoided crossing. The peak upconversion (\textbf{b}) and downconversion (\textbf{c}) occur at the mechanical frequency. \textbf{d.} We extract the bidirectional conversion efficiency from the upconversion and downconversion measurements as well as a microwave transmission and optical reflection measurement.}
	\label{SI_figure4}
\end{figure}

In addition to the many electromechanically active modes, there are also multiple optomechanicallly active modes. We can use another of these modes to operate the transducer at a different frequency, $\omega\und{m}/2\pi = 5.072$~GHz. We measure the CW conversion of the transducer with a 4-port measurement using the same techniques for the other transduction mode. We find that this mode is less strongly coupled with a smaller $g\und{em}/2\pi = 1.76 \pm 0.10$~MHz. The mechanical linewidth of this mode is also smaller with $\kappa\und{m}/2\pi = 0.54 \pm 0.01$~MHz. The conversion again displays an avoided crossing between the microwave and mechanical modes with a peak efficiency of 0.5\% with 1 $\upmu$W input power (see Figure~\ref{SI_figure4}).

\end{document}